\documentclass[conference]{IEEEtran}
\IEEEoverridecommandlockouts
\usepackage{booktabs}
\usepackage{multirow}
\usepackage{textcomp}
\usepackage{xcolor}
\usepackage{cite}
\usepackage{latexsym}
\usepackage{graphicx}
\usepackage{subcaption}
\usepackage{amsfonts,amssymb,amsmath}
\usepackage{nccmath}
\usepackage{optidef}
\usepackage{hyperref}
\usepackage{color,soul}
\usepackage{array}
\usepackage{algorithm}
\usepackage{algorithmic}
\usepackage[T1]{fontenc}
\usepackage[utf8]{inputenc}
\usepackage{fancyhdr}
\usepackage{lastpage}
\usepackage{caption}
\usepackage{soul}
\usepackage{subcaption}

\usepackage{amssymb,color}
\usepackage{amsmath}
\usepackage{bbm}
\usepackage{graphicx}
\usepackage{epstopdf}
\usepackage{amsthm,amsfonts}
\usepackage{tikz}

\usepackage{subcaption}

\usepackage{hhline}
\usepackage{multirow}

\usepackage{accents}

\usepackage{algorithm}
\usepackage{algorithmic}
\usepackage{psfrag}
\usepackage{cite}
\usepackage{pifont}
\usepackage[font={small}]{caption}

\allowdisplaybreaks

\usepackage{setspace}
\usepackage{bm}
\def\BibTeX{{\rm B\kern-.05em{\sc i\kern-.025em b}\kern-.08em
    T\kern-.1667em\lower.7ex\hbox{E}\kern-.125emX}}
\begin{document}
\title{Successive Interference Cancellation for ISAC in a Large Full-Duplex Cellular Network
}
\author{\IEEEauthorblockN{Konpal Shaukat Ali{$^\dagger$}, Roberto Bomfin{$^\dagger$} and Marwa Chafii{$^{\dagger *}$}}
\IEEEauthorblockA{$^{\dagger}$ Engineering Division, New York University (NYU) Abu Dhabi, 129188, UAE (Email: \{konpal.ali, roberto.bomfin\}@nyu.edu).\\ $^{*}$ NYU WIRELESS, NYU Tandon School of Engineering, Brooklyn, 11201, NY (marwa.chafii@nyu.edu).}
}
\maketitle

%75- to 100-word WCL abstract limit
%75- to 200-word abstract TCOM  JSAC FD
%wcnc edas allows 250
%currently 232
\begin{abstract}
To reuse the scarce spectrum efficiently, a large full-duplex cellular network with integrated sensing and communication (ISAC) is studied. Monostatic detection at the base station (BS) is considered. At the BS, we receive two signals: the communication-mode uplink signal to be decoded and the radar-mode signal to be detected. After self-interference cancellation (SIC), inspired by NOMA, successive interference cancellation (SuIC) is a natural strategy at the BS to retrieve both signals. However, the ordering of SuIC, usually based on some measure of channel strength, is not clear as the radar-mode target is unknown. The detection signal suffers a double path-loss making it vulnerable, but the uplink signal to be decoded originates at a user which has much lower power than the BS making it weak as well. Further, the intercell interference from a large network reduces the channel disparity between the two signals. We investigate the impact of both SuIC orders at the BS, i.e., decoding $1^{st}$ or detecting $1^{st}$ and highlight the importance of careful order selection. We find the existence of a threshold target distance before which detecting $1^{st}$ is superior and decoding $2^{nd}$ does not suffer much. After this distance, both decoding $1^{st}$ and detecting $2^{nd}$ is superior. Similarly, a threshold UE power exists after which the optimum SuIC order changes. We consider imperfections in SIC; this helps highlight the vulnerability of the decoding and detection in the setup.

\end{abstract}
%\begin{IEEEkeywords}
%Integrated sensing and communication (ISAC), stochastic geometry, full-duplex, successive interference cancellation, decoding order
%\end{IEEEkeywords}

\section{Introduction}
Spectrum is a scarce resource and we are constantly in a quest of new ways to reuse it efficiently. With growing demands in sensing, in addition to those in communication, for next generation networks, integrated sensing and communication (ISAC) has been proposed as a solution \cite{jrc_overview0}. With ISAC, we are able to use a single signal for both communication and radar-mode sensing \cite{12challenges}. This improves spectral efficiency, reduces power consumption and also reduces network interference \cite{myISACglobecom,jrc_sg4}. Full-duplex (FD) communication has also been proposed as a very promising solution to the spectrum scarcity problem \cite{FD8}. FD allows a node to both transmit and receive simultaneously in the same frequency channel. While FD does not double throughput, due to the presence of network interference and imperfections in self-interference cancellation (SIC), very promising gains have been shown \cite{myD2DFD}.

%Most works on ISAC focus on a setup with a single/few cells \cite{jrc_irs1,Bazzi_JRC1,Bazzi_JRC2}

Most works on ISAC focus on a setup with a single/few cells. Since real networks are becoming very dense and experience high interference, studying setups with a single/few cells can lead to inaccurate results and conclusions \cite{my_nomaMag}. Studying large networks that accurately model real-world networks is thus important. Stochastic geometry provides a unified mathematical paradigm for modeling large wireless networks and characterizing their operation while taking intercell interference into account \cite{jrc_sg4,myNOMA_tcom}. Works such as \cite{jrc_sg5,myISACglobecom} study ISAC in large networks; however, they study half-duplex networks where the BS only receives the radar-mode reflection signal. 

In this work, we study ISAC in a large FD cellular network with monostatic detection at the BS. This way, the spectrum is utilized very efficiently as the BS and UEs simultaneously transmit and receive communication-mode signals in the same frequency channel; further, radar-mode detection also happens at the same time on the same frequency channel, making use of the downlink signal for probing targets. %At the BS, we receive both the radar-mode reflection to be detected and the uplink communication-mode signal to be decoded.

%\textcolor{magenta}{The works in \cite{jrc_sg5,myISACglobecom} have studied ISAC in a large network taking into account intercell interference. Energy efficiency of ISAC is studied in \cite{jrc_sg5}, and a novel dynamic transmission strategy for ISAC that trades quantity of radar detection for quality was proposed in \cite{myISACglobecom}. The aforementioned works focus on half-duplex cellular networks. In this work, we are interested in studying ISAC in a large FD cellular network. This way, the BS and UEs simultanouesly transmit and receive communication-mode signals in the same frequency channel; further, radar-mode detection also happens at the same time on the same frequency channel, making use of the downlink signal for probing targets. }

ISAC in FD cellular networks has been studied in \cite{ISAC_FDcell1,ISAC_FDcell2,ISAC_FDcell3,ISAC_FDcell4}, but all of these works consider single-cell setups and do not take into account intercell interference. In \cite{ISAC_FDcell1,ISAC_FDcell2,ISAC_FDcell4} the BS transmits a downlink signal that is used for both communication and probing the target; in \cite{ISAC_FDcell3} it is not used for communication but only for probing. In all of these works, the communication-mode uplink signal and radar-mode reflection from the target are received together at the BS; the BS first decodes and removes the communication message and then detects the radar-mode target.

%an uplink communication-mode signal as well as the radar-mode echo
In a FD cellular network employing ISAC with monostatic detection, {different from other ISAC and FD setups,} we receive two signals of interest at the BS: the communication-mode uplink \emph{to be decoded} and the radar-mode reflection \emph{to be detected}. Inspired by NOMA, successive interference cancellation (SuIC) is a natural strategy in this scenario. SuIC requires ordering the signals based on some measure of channel strength \cite{myNOMA_tcom}. However, since the target is unknown, the channel strength is unknown and ordering is not straightforward. The works in \cite{ISAC_FDcell1,ISAC_FDcell2,ISAC_FDcell3,ISAC_FDcell4} assume that the communication-mode signal is stronger, decode first and then detect; however, this may not the be most accurate order. This is because: 1) while the radar-mode signal, sent from the BS to the target and back to the BS, suffers a double path loss making it more vulnerable, the communication-mode uplink signal is also weak as it originates at the UE which typically has lower transmit power than the BS, 2) taking into account intercell interference from all of the transmitting BSs and UEs in a large FD network significantly impacts both the radar and communication modes reducing disparity in their performance. 

Different from the aforementioned works, we study a large FD cellular network with ISAC using monostatic detection at the BS and taking into account intercell interference. We do not assume the order of SuIC at the BS and study both the scenario where decoding is done first followed by detection, as well as the new approach of detecting first and then decoding. To the best of our knowledge, this is the first work to: 1) account for intercell interference in a FD cellular network with ISAC, 2) study the impact of SuIC ordering of detection and decoding at the BS. Our contributions can be summarized as:
\begin{itemize}
\item We provide a tractable analytical framework {taking into account the impact of interfering BSs and UEs more precisely than previous works at both receivers of interest.}%
\item We quantify the probabilities of radar-mode detection and communication-mode decoding in both the downlink and in the uplink for both orders of SuIC at the BS.
\item We find the existence of of a superior SuIC order depending on whether targets inside or outside of a certain threshold distance are being searched for (e.g. within or outside of a boundary wall). We also find that the optimum SuIC order changes when the UE's transmit power exceeds a threshold. We show that if intercell interference is not considered, it appears that decoding first is always the superior order, which is misleading.
\item As the cellular network operates in FD, we highlight the vulnerability of decoding at the UE, and decoding and detection at the BS for both orders, due to imperfections in SIC. We show that after certain levels of residual self interference (RSI) decoding/detection is not viable and highlight the sensitivity of each mode.
\end{itemize}

The paper is organized as follows: Section II describes the system model. The methodology and detection strategies are in Section III. The SINR analysis is in Section IV. Section V presents the results and Section VI
concludes the paper.

\textit{Notation:} We denote vectors using bold text, $\|\textbf{z}\|$ is used to denote the Euclidean norm of the vector $\textbf{z}$. The ordinary hypergeometric function is denoted by ${}_2F_1$. The cdf (ccdf, pdf) of the random variable (RV) $X$ is denoted by $F_X$ ($\bar{F}_X$, $f_X$). The Laplace transform (LT) of the RV $X$ is denoted by $\mathcal{L}_X(s)=\mathbb{E}[e^{-sX}]$.

\section{System Model}\label{SysMod}

\begin{figure}[t]
\begin{minipage}[t]{\linewidth}
\centering\includegraphics[width=0.45\linewidth]{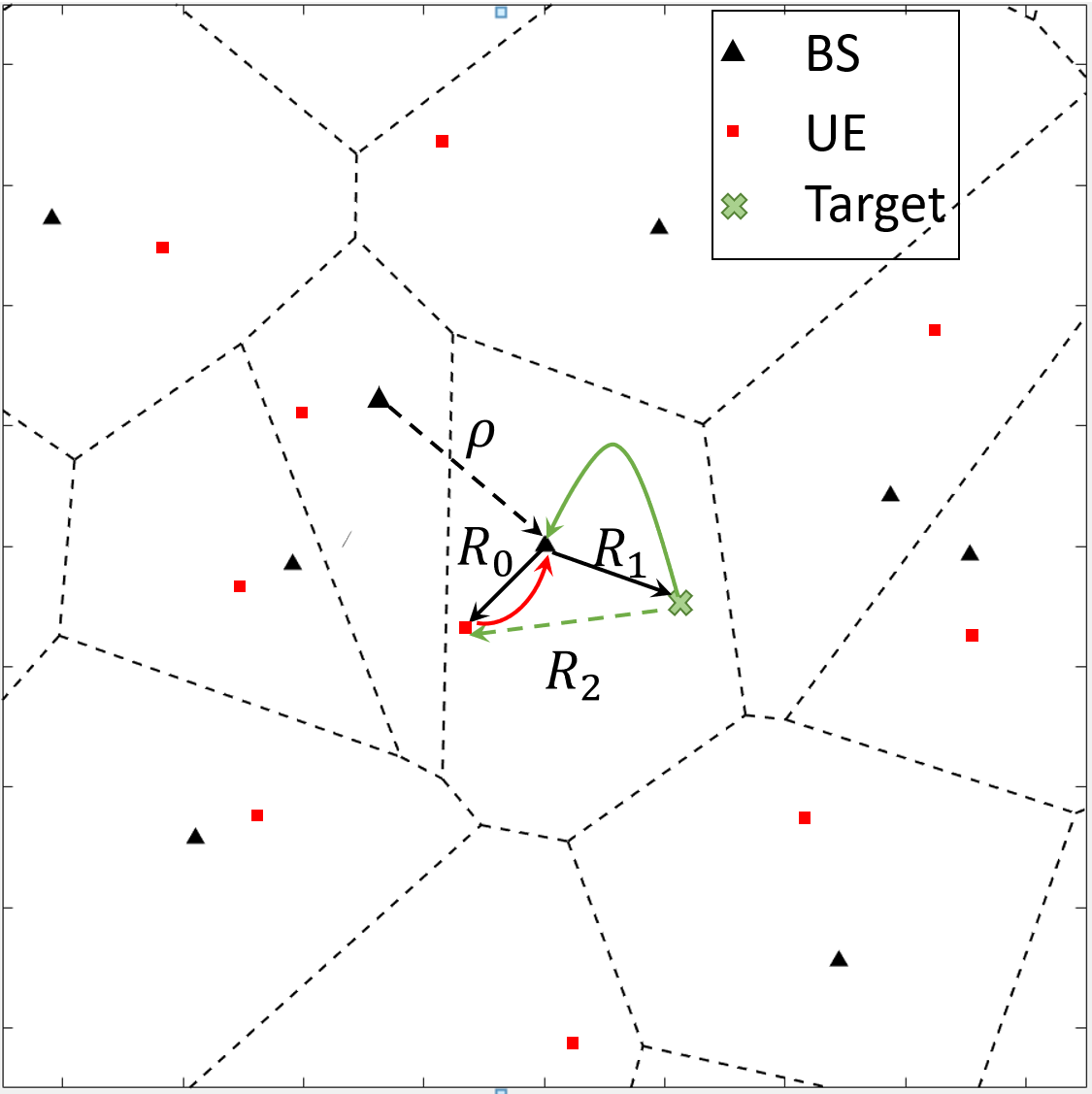}
\end{minipage}%\;\;\;
%\begin{minipage}[htb]{0.47\linewidth}
%\centering\includegraphics[width=0.9\linewidth]{figs/FD_FRad.png}
%\subcaption{Monostatic radar-only network}\label{frameworkR}
%\end{minipage}
\caption{A snapshot realization of the network. All the nodes of interest in the typical cell have been shown. Solid (dashed) arrows show signals of interest (interfering signals). The color of the link distance matches it arrow.}\label{framework5}
%\vspace{-0.1cm}
\end{figure}

\subsection{Network Model}

%which incorporates a weak repulsion from not only the BSs in the network but also the UEs in other cells

We consider a cellular network where BSs are distributed according to a homogeneous Poisson point process (PPP) $\Phi_{\rm B}$ with intensity $\lambda$. The communication-mode UEs are distributed uniformly and independently at random in the Voronoi cells of the BSs. We use $\Phi_{\rm U}$ to denote the point process (PP) of the UEs. As $\Phi_{\rm U}$ is a perturbed version of $\Phi_{\rm B}$ with each point perturbed uniformly at random inside its own Voronoi cell, $\Phi_{\rm U}$ is the user model of type I in \cite{mh_ue_PP}. For tracability $\Phi_{\rm U}$ is assumed to be independent of $\Phi_{\rm B}$. Each UE in a cell is assigned a unique frequency channel in a time slot to avoid intracell interference among the communication-mode UEs in that cell. We study one such frequency channel. The cellular network operates in FD thus the cellular UE and serving BS transmit in the same frequency channel and time slot and use SIC to cancel the interference of the message they are transmitting at their receive antenna.

The network under consideration has ISAC capability; in addition to the communication downlink and uplink operating in FD, it reuses the spectrum for radar-mode detection as well. This is done by using the downlink communication signal of the BS to probe and sense the target at an unknown location inside the cell. Thus, using ISAC, additional time, frequency and power resources are not spent for detection in the radar-mode. Targets to be detected are present in the network. The BS detects the reflection from the target; as the BS is the source of the probing signal, this detection is monostatic. Since the cellular network operates in FD, there are two signals of interest received at the BS: 1) the cellular uplink message from the UE \emph{to be decoded}, 2) the reflection/echo from the target \emph{to be detected}. As these two signals arrive at the BS in the same frequency channel and time slot, similar to NOMA, the BS uses SuIC to decode/detect these messages. Thus, the BS uses SIC to remove its own message and then SuIC for detection and decoding of the uplink. {We take into account imperfections is SIC via leakage of a fraction $\zeta$ of the self interference referred to as residual self interference (RSI).}

%Using coordination between the radars and BS, we also have the ability

To the network we add a BS at the origin $\textbf{o}$, which, under expectation over $\Phi_{\rm B}$, becomes the typical BS (tBS) communicating with the typical UE (tUE) which is distributed uniformly at random in the typical cell. The echo of the tBS's message from the typical target (tTar) is listened for at the tBS. In the remainder of this work, we study the performance of the typical cell. Since $\Phi_{\rm B}$ does not include the BS at $\textbf{o}$, the set of interfering BSs contributing to the intercell interference at the tUE is $\Phi_{\rm B}$. As the network is also operating in FD, the set of interfering UEs, $\Phi_{\rm U}$, also contribute to the intercell interference at the tUE. {At the tBS, intercell interference is received from the set of interfering BSs $\Phi_{\rm B}$ and from the set of interfering UEs $\Phi_{\rm U}$.} Fig. \ref{framework5} is a snapshot of the ISAC network where the nodes in the typical cell are shown. The tBS lies a distance $R_0$ from the tUE and $R_1$ from the tTar. The distance between the tTar and the tUE is denoted by $R_2$. The tBS is a distance $\rho$ from its nearest interferer.

\subsection{Channel Model and Link Distance Distribution}
We assume Rayleigh fading such that the fading coefficient between any two nodes is i.i.d. with a unit mean exponential distribution. A power-law path loss model is considered where the signal decays at the rate $r^{-\eta}$ with distance $r$, $\eta>2$ denotes the path loss exponent and $\delta=\frac{2}{\eta}$. BSs transmit with power $P_b$ and UEs transmit with lower power $P_u$.

%Similarly, in the radar-only network, the targets are distributed uniformly and randomly in the cells of radars distributed according to a PPP.

%\subsubsection{ISAC Network} 
As the UEs are distributed uniformly and randomly in the Voronoi cells of PPP distributed BSs with intensity $\lambda$, the distribution of the link distance $R_0$ follows $f_{R_0} (x)=2\pi b \lambda x \exp(-\pi b \lambda x^2)$, $x \geq 0,$
%\begin{align}
%f_{R_0} (x)=2\pi b \lambda x \exp(-\pi b \lambda x^2), \;\;\;\; x \geq 0, \label{f_R0}
%\end{align}
where $b=13/10$ is the correction factor due to the fact that the nodes of interest are in the typical cell, not in the 0-cell \cite[(12)]{mh_ue_PP}. %\textcolor{green}{Note that the location of the target, which is being detected, is unknown; however, since targets are also assumed to be distributed uniformly form a PPP, we know the distribution of the link distance $R_1$.} 
%In the ISAC network, 
We assume that the distance $R_1$ between the tBS and tTar is fixed. {Due to the PPP nature of the interferers, the distance of the tBS to its nearest interferer, $\rho$, follows $f_{\rho}(r)=2\pi \lambda r \exp(-\pi \lambda r^2)$, $r \geq 0.$ 
%\begin{align}
%f_{\rho}(r)=2\pi \lambda r \exp(-\pi \lambda r^2), \;\;\;\; r \geq 0. \label{f_rho}
%\end{align}}% and the tTar lies in a random direction. \textcolor{cyan}{The angle between the tBS-tTar link and the tTar-tUE link, denoted by $\phi$ (cf. Fig. \ref{framework5}), follows the distribution $f_{\phi}(u) =1/(2\pi)$, $0\leq u \leq 2\pi$.} \textcolor{green}{As a result, the distance between the tRad and tBS at $\textbf{o}$, denoted by $R_3$, is
%\begin{align}
%R_3=\sqrt{R_1^2 + r_2^2 - 2 r_2 R_1 \cos \phi}. \label{R3}
%\end{align}}
{At the tUE we are interested in the signal coming from the tBS. This signal undergoes fading; in this work, Rayleigh fading is used to model the infinite reflections coming from different scatterers in the environment. The reflection from the tTar ($R_2$ away from the tUE) is one of the infinite reflections received at the tUE. We thus do not need to take into account the reflection from the tUE separately\footnote{Taking into account this reflection separately in the simulation does not alter the results either, not shown for brevity, again justifying our rationale.}. {The statistics of $R_2$ are thus not required.}}

\section{Methodology and Detection Strategy }

%Transmission and Detection Strategies and Analysis Methodology

\subsection{Methodology of Analysis}
Fixed-rate transmissions are used in this work, such that a message is sent with transmission rate $\log(1+\theta_i)$ corresponding to an SINR threshold of $\theta_i$, where $i=b$ ($i=u$) for the downlink (uplink) message sent from the tBS (tUE). Such transmissions result in a throughput that is lower than the transmission rate because of possible outages. The performance of the communication-mode is typically measured using the decoding/coverage probability (i.e., the SINR ccdf), which is a measure of reliability, and throughput. Radar-mode performance was traditionally analyzed using metrics like the detection probability and faulty error probability. More recently, however, works on ISAC such as \cite{jrc_sg5,jrc_sg4} and even radar only \cite{sg_radar3} have switched to analyzing the performance of the radar-mode using the statistics of the SNR or SINR. This is due to the ability of such metrics to offer physics based insights into performance of the system as well as tractability of their statistics \cite{jrc_sg4} which allow us to draw meaningful conclusions, {highlight optimum SuIC ordering in this work} and shed light on careful parameter selection to attain different network objectives. In our work, we also focus on analyzing radar-mode performance using the statistics of the SINR at the BS that we refer to as the detection probability.%, as a measure of reliable detection. 

\subsection{SuIC at the BS}
Since the network under consideration is a FD cellular network, with ISAC, the tBS receives both the communication uplink signal from the tUE as well the radar-mode echo from the tTar. The BS thus needs to do both decoding of the communication uplink signal and detection of the radar-mode target. Inspired by NOMA, a natural approach to this issue is SuIC decoding where the tBS first decodes and removes the stronger signal and then decodes the weaker signal. In the case of uplink NOMA, the decoding order is typically established based on some measure of channel strength such as the distance of the UEs from the BS. In the case of a FD cellular network with ISAC, the situation is more complicated because the distance between the BS and target is unknown. {The detection signal of the radar-mode received at the BS undergoes a double path loss due to the signal first going from the BS to the target and then back. This significantly weakens the signal, particularly in comparison to the downlink communication-mode signal. While this may elude that the detection signal is lower than the uplink communication-mode signal too, this may not necessarily be the case as the transmit power of the BS is typically much larger than that of the cellular UE. Thus, while the detection signal suffers a double path loss, it may not necessarily be weaker than the uplink signal. Further, intercell interference from a large network reduces the disparity in SINR of the two signals. }

%Further, while the detection signal suffers a double path loss, we cannot assume that at the BS the detection signal received from the target will necessarily be much weaker than the decoding signal from the UE as the transmit power of the UE, $P_u$, is much lower than the transmit power of the BS.

In light of this, the superior order for SuIC is not clear. We therefore analyze the statistics of the SINR at the tBS for both the scenario where the signal from the tUE is detected first and when the signal from the tTar is decoded first. In each case, the signal is then removed and the other signal is decoded/detected after. Our results will shed light on the superior order under different circumstances and the significance of using the optimum order. As SuIC is a joint event, the success of the second signal decoded/detected depends on successful detection/decoding of the first signal. This is taken into account in the mathematical analysis in the following section.

\section{SINR Analysis}
\subsection{Statistics of Fading and Intercell Interference}
The fading coefficient between the tBS at \textbf{o} and the tUE at \textbf{u} (tTar at \textbf{t}) is $h_0$ ($h_1$). Due to Rayleigh fading, $h_0$ and $h_1$ are independent unit-mean exponential RVs. The radar-mode signal goes from the tBS to the tTar and then back to the tBS; the joint fading experienced at the tBS from the radar-mode is $h_{j_r}=h_1^2$. Obtaining the exact statistics of the joint fading RV $h_{j_r}$ is straightforward, however, this distribution does not lead to tractability in obtaining the SINR statistics for the radar-mode signal received at the tBS. Instead, we make use of the tight approximation for the statistics of $h_{j_r}$ proposed in \cite{myISACglobecom} that allows tractability in obtaining the SINR statistics in our setup.
The cdf of $h_{j_r}$, with $\mathbb{E}[h_{j_r}]=2$, is approximated as
\begin{align}
F_{h_{j_r}}(x) = \left(1-\exp \left(-\epsilon_r y \right)\right)^{m_r}, \label{hjr_cdf}
\end{align}
where $m_r=\sqrt{3/20}$ and $\epsilon_r=\text{harmonic}(m_r)/2$. 

%\textbf{\emph{Proof:}} Along the lines of {the proof of} Proposition 2 and using the fact that $\mathbb{E}[h_{j_r}]=\mathbb{E}[{h_1^2}]=\text{Var}[h_1]+(\mathbb{E}[h_1])^2=2$ as $h_1\sim \exp(1)$, we obtain \eqref{hjr_cdf}. In Fig. \ref{cdf_hj} we observe that the exact distribution based on simulations closely matches the analytical approximate in \eqref{hjr_cdf}. \qed

%\begin{figure}
%\begin{minipage}[htb]{0.98\linewidth}
%\centering\includegraphics[width=0.8\textwidth]
%{figs/cdf_hJ_both.eps}
%\caption{The simulated and proposed analytical cdf of $h_j$ and $h_{j_r}$ in \eqref{hj_cdf} and \eqref{hjr_cdf}, respectively.}\label{cdf_hj}
%\end{minipage}
%\end{figure}

%\subsection{Statistics of the Intercell Interference}

We denote the receivers of interest using $z=u$ for the the tUE located at $\textbf{u}$ and $z=o$ for the the tBS located at $\textbf{o}$. 
%Let $\nu \in \{\rm{B}, \rm{U} \}$ denote the source of interfering BSs and UEs, respectively, in $\Phi_\nu$.
The intercell interference from the PP of interfering BSs (UEs) $\Phi_{\rm B}$ ($\Phi_{\rm U}$) experienced at $z$ scaled to unit transmission power is denoted by $I_z^{\rm B}$ ($I_z^{\rm U}$). Let $\nu \in \{ \rm{B}, \rm{U} \}$ denote the source of interferers in $\Phi_\nu$. The intercell interference is $I_z^{\nu}=\sum_{\textbf{x} \in \Phi_{\nu}}  g_{\textbf{x}_z} {\|\textbf{x}-\textbf{z}\|}^{-\eta}$,
%\begin{align}
%I_z^{\nu}=\sum_{\textbf{x} \in \Phi_{\nu}}  g_{\textbf{x}_z} {\|\textbf{x}-\textbf{z}\|}^{-\eta}, \label{interf}
%\end{align} 
where $g_{\textbf{x}_z}$ denotes the fading coefficient from the interferer at $\textbf{x}$ to the receiver located at $\textbf{z}$.

 Let $\psi$ denote the guard zone distance, which is the radius of the disk centered at receiver $z$ inside which an interferer cannot exist. In some scenarios, there is an interferer guaranteed to be at the boundary of the guard zone, i.e., an interferer is conditioned to exist at a distance $\psi$ from receiver $z$. When an interferer is not conditioned to exist at the guard zone boundary, the LT of $I_z^\nu$, conditioned on the distance $\psi$, is
\begin{align}
\mathcal{L}_{I_z^\nu \mid \psi}^{\rm unc}(s)&=\exp \left({   \frac{-2 \pi \lambda s }{(\eta-2){\psi}^{\eta-2}} { }_2F_1  \left( 1,1  \!-\!  \delta; 2\! - \! \delta; \frac{-s}{{\psi}^{\eta}} \right)   }\right)     \label{L_I_uncond}\\
&\stackrel{\eta=4}= \exp \left(-\pi \lambda \sqrt{s} \tan^{-1} \left({\sqrt{s}}{\psi^{-2}} \right) \right) .
\end{align}
We obtain \eqref{L_I_uncond} by using the probability generating functional (pgfl) of the homogeneous PPP with intensity $\lambda$, the guard zone distance ${\psi}$ and $g_{\textbf{x}_z} \sim \exp(1)$ due to Rayleigh fading.% \cite{mhaenggi_Book}. 

When the nearest interferer is conditioned to exist a distance $\psi$ away from $z$, the LT $I_z^\nu$, conditioned on the distance $\psi$, is
{\small
\begin{align}
\!\mathcal{L}_{I_z^\nu \mid \psi}^{\rm cnd}(s ) \! &=\!\exp \!\left( \! {   \frac{-2 \pi \lambda s }{(\eta-2){\psi}^{\eta-2}} { }_2F_1 \! \left(\! 1,1 \!  - \!  \delta; 2 - \delta; \frac{-s}{{\psi}^{\eta}} \! \right) \!   }\right) \! \frac{1}{1+s \psi^{-\eta}}     \label{L_I_cond}\\
&\stackrel{\eta=4}= \exp \left(-\pi \lambda \sqrt{s} \tan^{-1} \left({\sqrt{s}}{\psi^{-2}} \right) \right) \frac{1}{1+s \psi^{-\eta}} .
\end{align}
}
The first term in \eqref{L_I_cond} is obtained along the lines of \eqref{L_I_uncond}. The second term comes from an interferer being guaranteed to exist at a distance $\psi$ from the receiver and as $g_{\textbf{x}_z} \sim \exp(1)$ \cite{myNOMA_tcom}.

At the tBS, the closest interfering BS is conditioned to exist at a distance $\rho$. Further, the closest interfering UE cannot be closer than $\rho/2$, which is the distance of the tBS to the closest cell edge; an interfering UE is not guaranteed to exist at this distance. At the tUE, the closest interfering BS cannot be closer than $R_0$, which is the distance to the serving BS. An interfering BS is not guaranteed to exist at this distance.  % and from $\Phi_{\rm U}$ is $\mathcal{L}_{I_u^{\rm U} }^{\rm rep}(s)$.

%At the tBS, the closest interfering BS is conditioned to exist at a distance $\rho$. Further, the closest interfering UE cannot be closer than $\rho/2$, which is the distance of the tBS to the closest cell edge; an interfering UE is not guaranteed to exist at this distance. Accordingly, at the tBS, the LTs of $I_o^{\rm B}$ and $I_o^{\rm U}$ are $\mathcal{L}_{I_o^{\rm B} \mid \rho}^{\rm cnd}(s)$ and $\mathcal{L}_{I_o^{\rm U} \mid \rho/2}^{\rm unc}(s)$, respectively. At the tUE, the closest interfering BS cannot be closer than $R_0$, which is the distance to the serving BS. An interfering BS is not guaranteed to exist at this distance. Thus, at the tUE, the LT of $I_u^{\rm B}$ is $\mathcal{L}_{I_u^{\rm B}\mid R_0 }^{\rm unc}(s)$. % and from $\Phi_{\rm U}$ is $\mathcal{L}_{I_u^{\rm U} }^{\rm rep}(s)$.

 %There is no minimum distance between the tUE and an interfering UE. However, the PP of interfering UEs exhibits some repulsion from the tUE. Accordingly, the LT of interference at the tBS and tUE from both PP of interferers is summarized in Table \ref{LT_Interf_Table}.

In the case of interference from the PP of UEs $\Phi_{\rm U}$ to the tUE there is no guard zone around the tUE. This is because the tUE can be close to a cell edge and an interfering UE can be close to that edge as well. However, there is a repulsion between the PP of interfering UEs, which form a user model of type I, and receivers at random location inside the typical cell \cite{mh_ue_PP} such as the tUE. At these receivers, $\Phi_{\rm U}$ appears as a soft-core process where the likelihood of having two points very close is lower than in a PPP. As seen at the tUE, $\Phi_{\rm U}$  is modeled as a non-homogeneous but isotropic PPP with radial intensity function $\lambda(r)=\lambda(1-\exp(-3\sqrt{\lambda}r))$ \cite[(8)]{mh_ue_PP}. The LT of $I_z^{\rm U}$ from $\Phi_{\rm U}$ at the tUE is %

{\small \begin{align}
\mathcal{L}_{I_u^{\rm U} }^{\rm rep}(s)&=\exp \left(\!   -2 \pi \lambda \int_0^{\infty}\! \frac{s r^{-\eta+1}}{1+ s r^{-\eta}} \left(\! 1-e^{-3\sqrt{\lambda}r} \!\right) dr    \! \right) . \label{L_I_repulsion}
\end{align} }
Using the pgfl of the PPP, the radial intensity function, and $g_{\textbf{x}_z} \sim \exp(1)$, we obtain \eqref{L_I_repulsion}.

The LT of interference at the tBS and tUE from both PP of interferers is summarized in Table \ref{LT_Interf_Table}.

\begin{table}[t]
  \centering
  \resizebox{\linewidth}{!}{\begin{tabular}{|c|c|c|c|}
    \hline
     \textbf{Receiver} & $z$
    &   \textbf{LT of Interference from $\Phi_{\rm B}$} & \textbf{LT of Interference from $\Phi_{\rm U}$} \\
    %\hhline{~--}
    \hline
    tBS  & $o$ & $\mathcal{L}_{I_o^{\rm B} \mid \rho}^{\rm cnd}(s)$ in \eqref{L_I_cond} & $\mathcal{L}_{I_o^{\rm U} \mid \rho/2}^{\rm unc}(s)$ in \eqref{L_I_uncond}   \\ \hline
    UE & $c$ & $\mathcal{L}_{I_u^{\rm B}\mid R_0 }^{\rm unc}(s)$ in \eqref{L_I_uncond} & $\mathcal{L}_{I_u^{\rm U} }^{\rm rep}(s)$ in \eqref{L_I_repulsion}  \\ \hline
  \end{tabular}}
  \caption{The LTs of interference, with transmit power scaled to 1, at the tBS and tUE in a FD cellular network with ISAC.}\label{LT_Interf_Table}
\end{table}

%\vspace{-0.5cm}

\subsection{Statistics of the SINR}
The SINR of the communication-mode downlink at the tUE in a FD cellular network with ISAC is
{\begin{align}
&{\rm SINR_u}= \frac{ P_b h_0 R_0^{-\eta}  }{ P_b I_u^{\rm B} + P_u I_u^{\rm U} + P_u \zeta +  \sigma^2 } , \label{SINR_u}
\end{align} 
where $P_b I_u^{\rm B}$ is the intercell interference from $\Phi_{\rm B}$, $P_u I_u^{\rm U}$ is the intercell interference from $\Phi_{\rm U}$, $P_u \zeta$ is the residual SI and $\sigma^2$ is the noise experienced at the tUE located at $\textbf{u}$. }% is $P_bI_u$.

%in a FD cellular network with ISAC
\textbf{\emph{Theorem 1:}} The probability of successful decoding of the downlink communication-mode message at the tUE is %The ccdf of the downlink communication-mode SINR at the tUE is
{\small \begin{align}
\!\!\mathbb{P}(\!{\rm SINR_u} \!>\!\theta_b \! )& \!=\!  \mathbb{E}_{\!R_0} \! \left[ \!   \mathcal{L}_{I_u^{\rm B} \mid \! R_0}^{\rm unc}  \!\left(\! \frac{ \theta_b  }{R_0^{-\eta}}  \!\right) \! \mathcal{L}_{I_u^{\rm U} }^{\rm rep}  \! \left( \! \frac{ \theta_b  P_u}{P_b R_0^{-\eta} }  \!\right) \! e^{ \frac{-\theta_b  (\sigma^2 \!+ \! P_u \zeta )}{P_b  R_0^{-\eta}}  }  \! \right] \!. \label{ccdfSINR_u}
\end{align} }
%{where $\mathcal{L}_{I_u^{\rm B} \mid R_{0}}(s)$ is given in \eqref{L_I_uncond}.} 

\textbf{\emph{Proof:}} Rewriting the SINR expression in \eqref{SINR_u} we obtain
{\small \begin{align*}
\mathbb{P}({\rm SINR_u}>\theta_b )=\mathbb{E}\left[ \bar{F}_{h_0} \left( \frac{\theta_b R_0^{\eta} }{ P_b} \left( P_b I_u^{\rm B} + P_u I_u^{\rm U} + P_u\zeta + \sigma^2 \right) \right) \right].
\end{align*}}
Since $h_0 \sim \exp(1)$, using the definition of the LT and the LTs of interference in Table \ref{LT_Interf_Table}, we obtain \eqref{ccdfSINR_u}. \qed

%As explained, the situation at the tBS is more complicated as both the communication-mode uplink needs to be decoded and the radar-mode signal needs to be detected. 

At the tBS, we study both the situations where: 1) the tBS $1^{st}$ decodes the uplink communication signal, removes it and then detects, 2) the tBS $1^{st}$ detects the radar detection signal, removes it and then decodes.

%At the tBS, we study both the situations where the tBS $1^{st}$: 1) decodes the uplink communication signal, removes it and then detects, 2) detects the radar detection signal, removes it and then decodes.

\subsubsection{Decode $1^{st}$ at BS}
The SINR of the communication-mode uplink at the tBS when decoding is done first is
{\begin{align}
&{\rm SINR_o^{\text{c-}1^{st}}}= \frac{  P_u h_0 R_0^{-\eta}  }{ P_b h_1^2 R_1^{-2\eta} + P_b I_o^{\rm B} + P_u I_o^{\rm U} +  P_b \zeta +  \sigma^2 } , \label{SINR_o_decode1}
\end{align} 
where $P_b h_1^2 R_1^{-2\eta}$ is the intracell interference from the detection signal, $P_b I_o^{\rm B}$ is the intercell interference from $\Phi_{\rm B}$, $P_u I_o^{\rm U}$ is the intercell interference from $\Phi_{\rm U}$, $P_b \zeta$ is the residual SI and $\sigma^2$ is the noise experienced at the tBS located at $\textbf{o}$. }% is $P_bI_u$.

After decoding and removing the uplink message, the SINR of the radar-mode signal at the tBS, where $h_1^2=h_{j_r}$, is
\begin{align}
{\rm SINR_{o, \text{c-}1^{st}}^{\text{r-}2^{nd}}}&= \frac{  P_b h_{j_r} R_1^{-2\eta}  }{   P_b I_o^{\rm B} + P_u I_o^{\rm U} +  P_b \zeta +  \sigma^2 }. \label{SINR_o_detect2}
\end{align}

\textbf{\emph{Theorem 2:}} The probability of successful decoding of the uplink communication-mode message at the tBS, when it is decoded first, is %The probability of successful decoding of the uplink message at the tBS when decoding is done first is % ccdf of the uplink communication-mode SINR at the tBS when decoding is done first is% the communication-mode signal is decoded first is
{\small \begin{align}
&\mathbb{P}( {\rm SINR_o^{\text{c-}1^{st}}} >\theta_u ) =  \mathbb{E}_{\rho,R_0} \Bigg[   \mathcal{L}_{I_o^{\rm B} \mid \rho}^{\rm cnd}  \left( \frac{ \theta_u R_0^{\eta} P_b}{P_u }  \right) \mathcal{L}_{I_u^{\rm U} \mid \rho/2 }^{\rm unc}  \left( {\theta_u R_0^{\eta}}  \right) \nonumber \\
& \times \frac{\sqrt{\pi P_u} e^{ \frac{-\theta_u  (\sigma^2 + P_b \zeta )}{P_u R_0^{-\eta}} } }{2\sqrt{ \!  {\theta_u R_0^{\eta} P_b  R_1^{-2\eta}}\! }} { \exp\left(\frac{1}{4} \frac{ P_u R_1^{2\eta}}{\theta_u R_0^{\eta} P_b  } \right) \text{erfc}\left( \sqrt{ \!  \frac{ P_u R_1^{2\eta}}{4\theta_u R_0^{\eta} P_b  } \! }\right)}   \Bigg]. \label{ccdfSINR_o_decode1}
\end{align}}
%{where $\mathcal{L}_{I_u^{\rm B} \mid R_{0}}(s)$ is given in \eqref{L_I_uncond}.} 

\textbf{\emph{Proof:}} Rewriting the SINR expression in \eqref{SINR_o_decode1} we have
{\footnotesize\begin{align*}
&\mathbb{P}(\! {\rm SINR_o^{\text{c-}1^{st}}}\!>\!\theta_u\! ) \!= \! %\mathbb{E} \! \left[ \! \bar{F}_{h_0} \left(\! \frac{\theta_u R_0^{\eta} }{ P_u} \left(\!  P_b I_o^{\rm B}  + P_u I_o^{\rm U}  + P_b\zeta + \sigma^2 + \frac{P_b h_1^2}{ R_1^{2\eta}} \! \right) \! \right) \! \right]\\
 { \mathbb{E} \!  \left[ \!    \frac{ \mathcal{L}_{I_o^{\rm B} \mid \rho}^{\rm cnd}  \!\left( \! \frac{ \theta_u  P_b}{P_u R_0^{-\eta} } \! \right)\! \mathcal{L}_{I_u^{\rm U} \mid \rho/2 }^{\rm unc}\!  \left( \! {\frac{\theta_u} {R_0^{\!-\eta}}} \! \right) \!  \mathcal{L}_{\!h{_1}^2} \!\left(\!  \frac{\theta_u R_0^{\eta} P_b  }{ P_u R_1^{2\eta}} \! \right)}{   \exp \left( \frac{\theta_u (\sigma^2 + P_b \zeta )}{P_u R_0^{-\eta} } \right)}  \! \right]}
\end{align*} }
using the cdf of $h_0 \sim \exp(1)$ and the LTs of interference in Table \ref{LT_Interf_Table}. {The LT of $h_1^2$ is derived as $\mathcal{L}_{h{_1}^2} (s)= \frac{\sqrt{\pi} \exp\left(\frac{1}{4s}\right) \text{erfc}\left(\frac{1}{2\sqrt{s}}\right)}{2\sqrt{s}}$;} applying this, we obtain \eqref{ccdfSINR_o_decode1}. \qed

\textbf{\emph{Lemma 1:}} The probability of successful detection of the radar-mode signal at the tBS, when the communication-mode uplink message has been decoded and removed, is % ccdf of the SINR of the radar-mode at the tBS, conditioned on the event that the communication-mode uplink signal has been decoded and removed is
{\small \begin{align}
%& \mathbb{P}( {\rm SINR_{o, \text{c-}1^{st}}^{\text{r-}2^{nd}}} \!> \! \theta  ) \!=\!  1 \!-\!  \mathbb{E}_{\rho} \! \left[ \!  \left( \!  1\! - \! \frac{ \mathcal{L}_{I_o^{\rm B} \mid \rho}^{\rm cond }   \left(\! \frac{\epsilon_r  \theta   }{R_1^{-2\eta}}  \!\right) \mathcal{L}_{I_o^{\rm U} \mid \rho/2 }^{\rm uncond }  \left(\! \frac{\epsilon_r  \theta   P_u }{P_b R_1^{-2\eta}}  \! \right)  }{\exp \! \left( \! {\epsilon_r  \theta R_1^{2\eta}  (\zeta + \sigma^2/P_b) }  \! \right)}  \!  \!  \right)^{m_r} \!  \right]. \label{ccdfSINR_o_detect2} %\\
& \mathbb{P}( {\rm SINR_{o, \text{c-}1^{st}}^{\text{r-}2^{nd}}} >\theta_b  )=  1 \!-\!  \mathbb{E}_{\rho} \Bigg[  \Bigg(  1\! - \! \mathcal{L}_{I_o^{\rm B} \mid \rho}^{\rm cond }   \left( {\epsilon_r  \theta_b R_1^{2\eta}  }  \right) \times \nonumber\\
& \mathcal{L}_{I_o^{\rm U} \mid \rho/2 }^{\rm uncond }  \left(\! \frac{\epsilon_r  \theta_b R_1^{2\eta}  P_u }{P_b}  \! \right)  \! \exp \! \left( \! \frac{-\epsilon  \theta_b R_1^{2\eta}  (P_b\zeta + \sigma^2) }{P_b}  \! \right) \!  \Bigg)^{m_r} \!  \Bigg]. \label{ccdfSINR_o_detect2}
\end{align} }
%{where $\mathcal{L}_{I_r^{\rm B} \mid \rho_r}(s)$ is obtained from \eqref{L_I_cond}.}

\textbf{\emph{Proof:}} Rewriting \eqref{SINR_o_detect2}, using $F_{h_{j_r}}$ in \eqref{hjr_cdf}, and applying the LTs of the interference in Table \ref{LT_Interf_Table}, we obtain \eqref{ccdfSINR_o_detect2}.  \qed

%\textbf{\emph{Proof:}} Rewriting the SINR expression in \eqref{SINR_o_detect2} we obtain
%\begin{align*}
%&\mathbb{P}({\rm SINR_{o, \text{c-}1^{st}}^{\text{r-}2^{nd}}} >\theta ) \!= \!\mathbb{E}\left[\! \bar{F}_{h_{j_r}} \left(\! \frac{\theta  \left(\! P_{\rm avg }I_r^{\rm B} + P_u I_r^{\rm U}+  P_b \zeta + \sigma^2 \!\right) }{ P_b R_1^{-2\eta}}  \! \right) \! \right].
%\end{align*} 
%Using the cdf of $h_{j_r}$ in \eqref{hjr_cdf} along with the definition of the LT and applying the LTs of the interference in Table \ref{LT_Interf_Table}, we obtain \eqref{ccdfSINR_o_detect2}.  \qed

%We denote the joint of event of successful detection of the radar-mode signal at the tBS after the communication-mode uplink signal has been decoded and removed as

\textbf{\emph{Corollary 1:}} Successful detection of the radar-mode signal at the tBS when the tBS detects $2^{nd}$ is the joint event of successful decoding $1^{st}$ (i.e., \eqref{ccdfSINR_o_decode1}) and successful detection after this (i.e., \eqref{ccdfSINR_o_detect2}). It is calculated as $\mathbb{P} \left( {\rm SINR_o^{\text{c-}1^{st}}} \!>\! \theta_u \right) \mathbb{P} \left( {\rm SINR_{o, \text{c-}1^{st}}^{\text{r-}2^{nd}}} \! > \! \theta_b \right)$.

\subsubsection{Detection $1^{st}$ at BS}
The SINR of the radar-mode detection signal at the tBS when detection is done first is
\begin{align}
&{\rm SINR_o^{\text{r-}1^{st}}}= \frac{ P_b h_{j_r} R_1^{-2\eta}    }{ P_u h_0 R_0^{-\eta} + P_b I_o^{\rm B} + P_u I_o^{\rm U} +  P_b \zeta +  \sigma^2 } , \label{SINR_o_detect1}
\end{align} 
where $P_u h_0 R_0^{-\eta}$ is the intracell interference from the decoding signal. % is $P_bI_u$.
\\After detecting and removing the radar-mode signal, the SINR of the communication-mode uplink message at the tBS is
\begin{align}
{\rm SINR_{o, \text{r-}1^{st}}^{\text{c-}2^{nd}}}&= \frac{  P_u h_0 R_0^{-\eta}  }{   P_b I_o^{\rm B} + P_u I_o^{\rm U} +  P_b \zeta +  \sigma^2 } . \label{SINR_o_decode2}
\end{align}

\textbf{\emph{Theorem 3:}} The probability of successful detection of the radar-mode signal at the tBS, when it is detected first, is% The SINR ccdf at the tBS when the radar-mode signal is detected first is
{\small \begin{align}
&\mathbb{P}( {\rm SINR_o^{\text{r-}1^{st}}} >\theta_b ) = 1 \!-\!  \mathbb{E}_{\rho} \Bigg[ \! \Bigg( \! 1\! - \! {\left( \! 1 \!+\! \epsilon_r \theta_b \frac{P_u R_1^{2\eta} }{P_b R_0^{\eta} } \! \right)^{-1}} \! \mathcal{L}_{I_o^{\rm B} \mid \rho}^{\rm cond } \!   \left(\! \frac{\epsilon_r  \theta_b}{R_1^{-2\eta}  } \! \right)  \nonumber\\
& \times \mathcal{L}_{I_o^{\rm U} \mid \rho/2 }^{\rm uncond }  \left(\! \frac{\epsilon_r  \theta_b R_1^{2\eta}  P_u }{P_b}  \! \right)   \exp \! \left( \! \frac{-\epsilon_r  \theta_b R_1^{2\eta}  (P_b\zeta + \sigma^2) }{P_b}  \! \right) \!   \Bigg)^{m_r} \!  \Bigg]. \label{ccdfSINR_o_detect1}
\end{align}}
%{where $\mathcal{L}_{I_u^{\rm B} \mid R_{0}}(s)$ is given in \eqref{L_I_uncond}.} 
\textbf{\emph{Proof:}} Rewriting \eqref{SINR_o_detect1}, using the cdf of $h_{j_r}$, $\mathcal{L}_{h_0}(s)$ and the LTs of interference in Table \ref{LT_Interf_Table}, we obtain \eqref{ccdfSINR_o_detect1}. \qed

\textbf{\emph{Lemma 2:}} The probability of successful decoding of the uplink communication-mode message at the tBS, after the radar-mode signal has been detected and removed, is   %The ccdf of the SINR of the communication-mode at the tBS, conditioned on the event that the radar-mode detection signal has been decoded and removed is
{\small \begin{align}
& \mathbb{P}( {\rm SINR_{o, \text{r-}1^{st}}^{\text{c-}2^{nd}}} >\theta_u  )=  \mathbb{E}_{\rho,R_0} \Bigg[   \mathcal{L}_{I_o^{\rm B} \mid \rho}^{\rm cnd}  \left( \frac{ \theta_u R_0^{\eta} P_b}{P_u}  \right) \mathcal{L}_{I_u^{\rm U} \mid \rho/2 }^{\rm unc}  \left( {\theta_u R_0^{\eta}}  \right) \nonumber \\
& \times \exp \left( \frac{-\theta_u R_0^{\eta}  (\sigma^2 + P_b \zeta )}{P_u } \right)  \Bigg].  \label{ccdfSINR_o_decode2}
\end{align} }
%{where $\mathcal{L}_{I_r^{\rm B} \mid \rho_r}(s)$ is obtained from \eqref{L_I_cond}.}

\textbf{\emph{Proof:}} Rewriting \eqref{SINR_o_decode2}, using the cdf of $h_0$ and applying the LTs of the interference in Table \ref{LT_Interf_Table}, we obtain \eqref{ccdfSINR_o_decode2}.  \qed

\textbf{\emph{Corollary 2:}} Successful decoding of the communication-mode uplink message at the tBS when the tBS decodes $2^{nd}$ is the joint event of successful detection $1^{st}$ (i.e., \eqref{ccdfSINR_o_detect1}) and successful decoding after this (i.e., \eqref{ccdfSINR_o_decode2}). It is calculated as $\mathbb{P} \left( {\rm SINR_o^{\text{r-}1^{st}}} \!>\! \theta_b \right) \mathbb{P} \left( {\rm SINR_{o, \text{r-}1^{st}}^{\text{c-}2^{nd}}} \! > \! \theta_u \right)$.

\textbf{\emph{Remark 1:}} When we refer to the probability of the BS detecting (decoding) $2^{nd}$ successfully, we are referring to the probability of the joint event in Corollary 1 (Corollary 2).

\section{Results}
In this section, unless mentioned otherwise, we consider BS intensity $\lambda=10^{-5}$, $\eta=4$, $P_u=0.2$, $P_b=1$ and $v=1/(60\sqrt{\lambda}) $. Simulations are repeated $10^4$ times. %\textcolor{green}{For fairness of comparison we also set $R_r=R_1$ and $P_r=P_l=1$. Since $R_r=R_1$ and $P_r=P_l$, in this section, the performance of the radar-only network (from \eqref{ccdfSINR_radOnly}) is identical to the performance of monostatic detection in the ISAC network without DTS (i.e., from \eqref{ccdfSINR_r_mo_noDTS}). This will be verified in Fig. \ref{verify} too.}

\begin{figure}[htb]
%\vspace{-0.4cm}
\begin{minipage}[htb]{0.484\linewidth}
\centering\includegraphics[width=\linewidth]{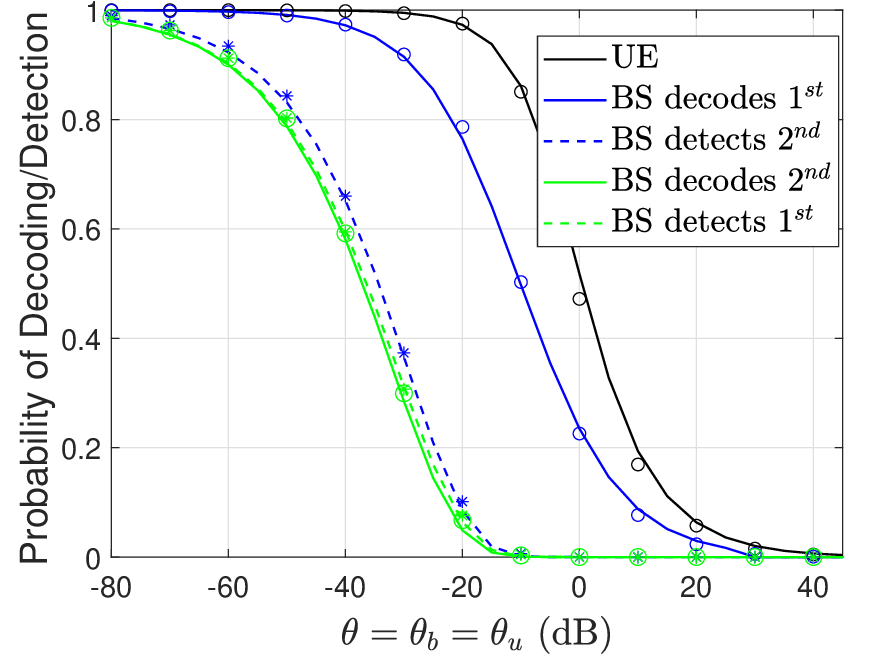}
\caption{The probability of detection or decoding vs. $\theta=\theta_b=\theta_u$; $R_1=5v$ and $\zeta=10^{-12}$. Lines (markers) represent the analysis (simulations).}\label{verify_conf}
\end{minipage}\;\;
\begin{minipage}[htb]{0.484\linewidth}
\centering\includegraphics[width=\linewidth]{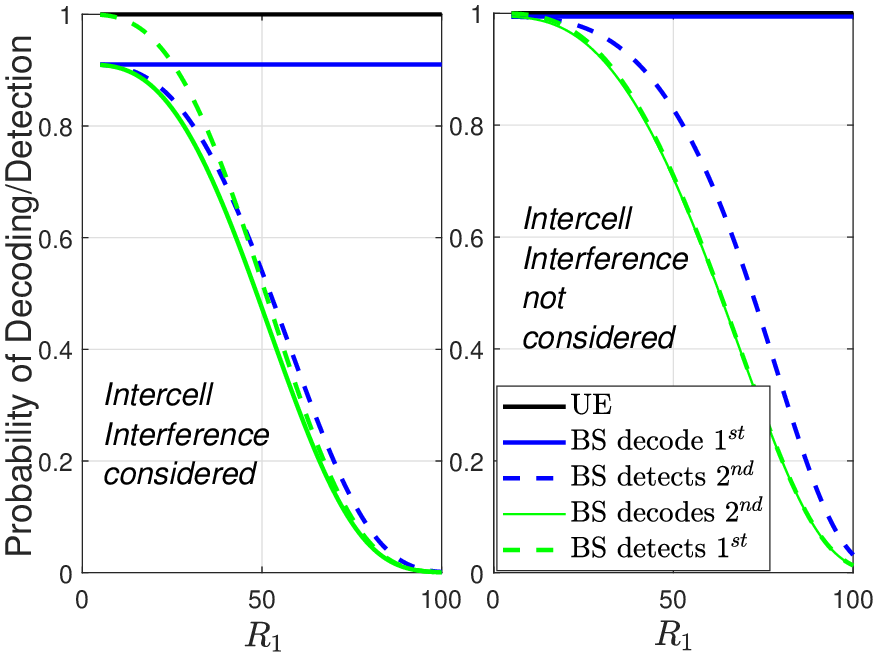}
\caption{The probability of detection or decoding vs. $R_1$. We use $\theta_u=-30$ dB, $\theta_b=-60$ dB and $\zeta=10^{-9}$.\\}\label{vsR1}
\end{minipage}
\vspace{-0.1cm}
\end{figure}

Fig. \ref{verify_conf} is a plot of the decoding and detection probabilities in a FD cellular network employing ISAC when $R_1=5v$ and $\zeta=10^{-12}$. The simulations match the analysis {verifying the accuracy of our mathematical analysis.} As anticipated, the probability of decoding at the UE significantly outperforms detection probabilities. This is due to the double path loss associated with the nature of detection. For the plotted scenario, decoding at the BS when it is done $1^{st}$ is better than when it is done $2^{nd}$ for both detection and decoding at the BS; however, as we will see, this does not hold in all circumstances. 

%\begin{figure}[thb]
%\begin{minipage}[htb]{\linewidth}
%\centering\includegraphics[width=0.64\linewidth]{figsF5/conf_vsR1andvsPu.eps}
%\caption{The probability of detection or decoding vs. $R_1$ and vs. $P_u$. We use $\theta_u=-30$ dB, $\theta_b=-60$ dB and $\zeta=10^{-9}$.}\label{vsR1andVsPu}
%\end{minipage}
%\begin{minipage}[htb]{\linewidth}
%\centering\includegraphics[width=0.64\linewidth]{figsF5/conf_vsR1andvsPu.eps}
%\caption{\textcolor{green}{The probability of detection or decoding vs. $R_1$ and vs. $P_u$. We use $\theta_u=-30$ dB, $\theta_b=-60$ dB and $\zeta=10^{-9}$.}}\label{vsR1andVsPu}
%\end{minipage}
%\begin{minipage}[htb]{\linewidth}
%\centering\includegraphics[width=0.64\linewidth]{figsF5/conf_vsPu_vsR1_detect.eps}
%\caption{The probability of successful detection or decoding vs. $R_1$ and $P_u$, using $\theta_u=-40$ dB, $\theta_b=-60$ dB and $\zeta=10^{-9}$.}\label{vsPuVsR1}
%\end{minipage}
%\begin{minipage}[htb]{\linewidth}
%\centering\includegraphics[width=0.64\linewidth]{figsF5/conf_vsZeta.eps}
%\caption{The probability of successful detection or decoding vs. $\zeta$ using $R_1=7v$, $\theta_u=\theta_b=-40$ dB.}\label{vsZeta}
%\end{minipage}
%%\vspace{-0.2cm}
%\end{figure}

\begin{figure*}[thb]
\begin{minipage}[htb]{0.3\linewidth}
\centering\includegraphics[width=0.76\linewidth]{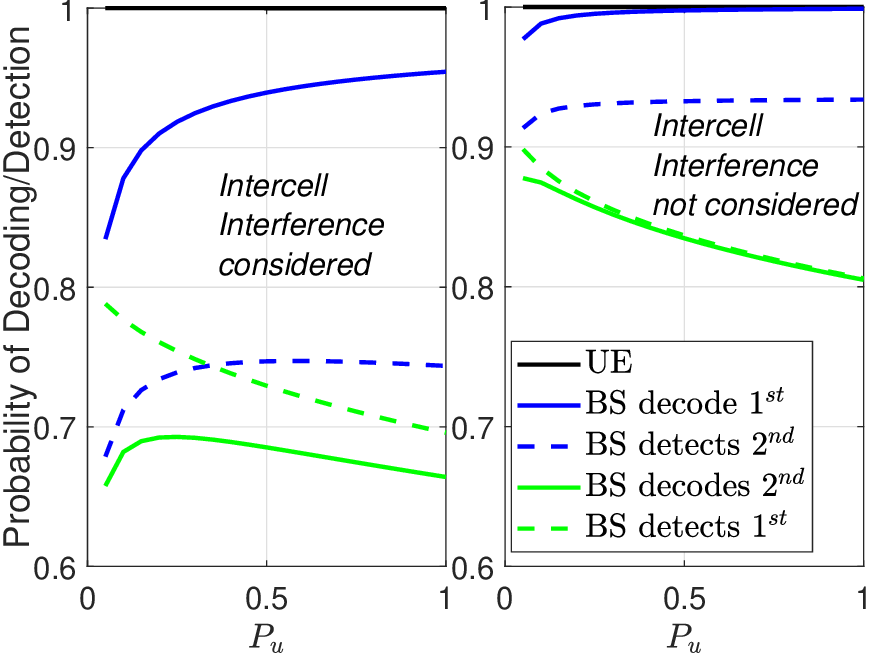}
\caption{The probability of detection or decoding vs. $P_u$. We use $R_1=7v$, $\theta_u=-30$ dB, $\theta_b=-60$ dB and $\zeta=10^{-9}$.}\label{vsPu}
\end{minipage}\;\;\;
%\begin{minipage}[htb]{0.3\linewidth}
%\centering\includegraphics[width=0.75\linewidth]{figsF5/conf_vsR1andvsPu.eps}
%\caption{The probability of detection or decoding vs. $R_1$ and vs. $P_u$. We use $\theta_u=-30$ dB, $\theta_b=-60$ dB and $\zeta=10^{-9}$.}\label{vsR1andVsPu}
%\end{minipage}\;\;\;
%\begin{minipage}[htb]{0.32\linewidth}
%\centering\includegraphics[width=\linewidth]{figsF5/conf_vsPu.eps}
%\caption{The probability of successful detection or decoding vs. $P_u$ using $R_1=7v$, $\theta_u=-30$ dB, $\theta_b=-60$ dB and $\zeta=10^{-9}$.}\label{vsPu}
%\end{minipage}\;\;\;
\begin{minipage}[htb]{0.3\linewidth}
\centering\includegraphics[width=0.76\linewidth]{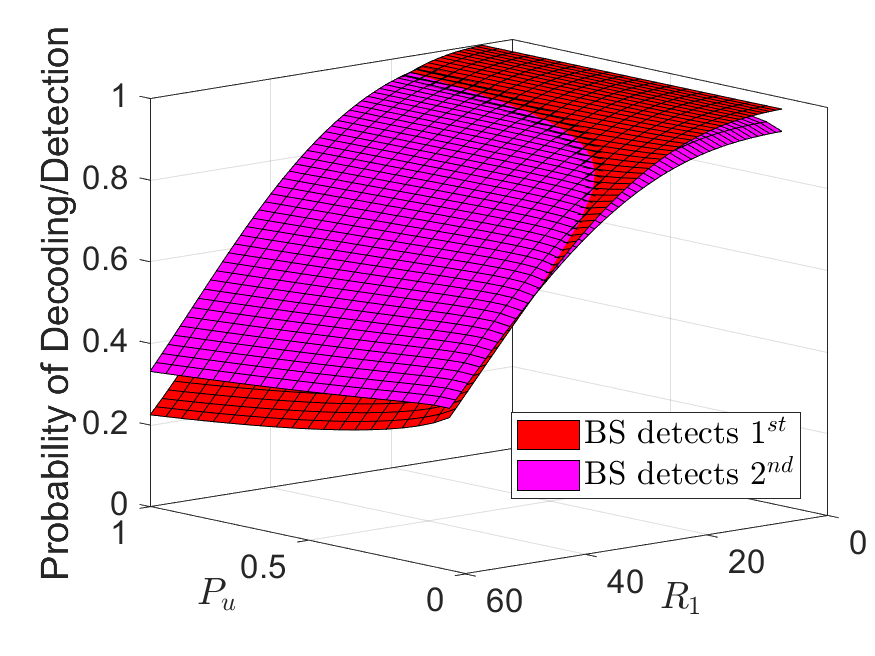}
\caption{The probability of detection or decoding vs. $R_1$ and $P_u$, using $\theta_u=-40$ dB, $\theta_b=-60$ dB and $\zeta=10^{-9}$.}\label{vsPuVsR1}
\end{minipage}\;\;\;
\begin{minipage}[htb]{0.3\linewidth}
\centering\includegraphics[width=0.76\linewidth]{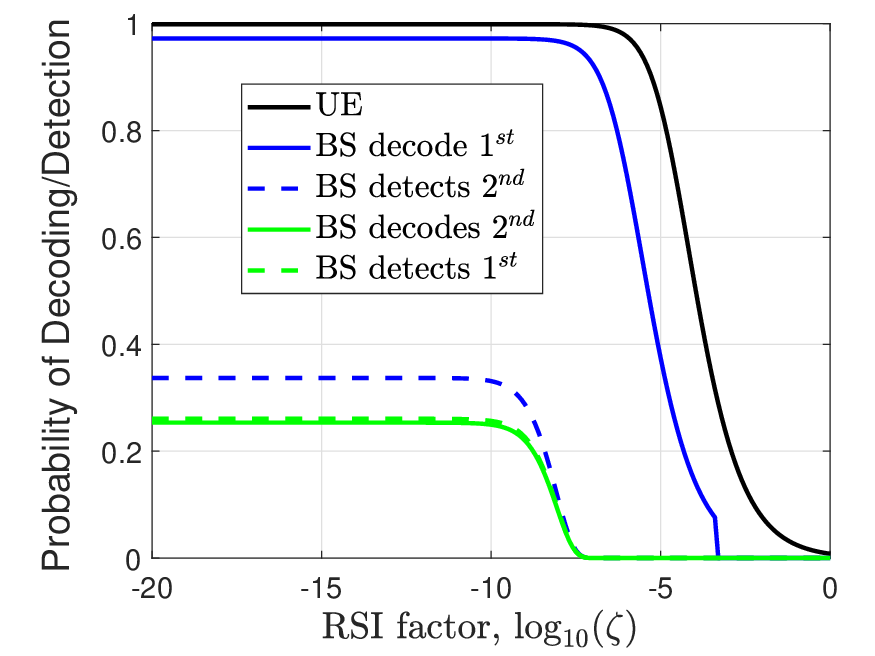}
\caption{The probability of detection or decoding vs. $\zeta$ using $R_1=7v$, $\theta_u=\theta_b=-40$ dB.\\}\label{vsZeta}
\end{minipage}
\vspace{-0.3cm}
\end{figure*}

%\begin{figure*}[thb]
%\begin{minipage}[htb]{0.24\linewidth}
%\centering\includegraphics[width=\linewidth]{figsF5/conf_vsR1andvsPu.eps}
%\caption{The probability of detection or decoding vs. $R_1$ and vs. $P_u$. We use $\theta_u=-30$ dB, $\theta_b=-60$ dB and $\zeta=10^{-9}$.}\label{vsR1andVsPu}
%\end{minipage}\;
%\begin{minipage}[htb]{0.24\linewidth}
%\centering\includegraphics[width=\linewidth]{figsF5/conf_vsR1andvsPu.eps}
%\caption{The probability of detection or decoding vs. $R_1$ and vs. $P_u$. We use $\theta_u=-30$ dB, $\theta_b=-60$ dB and $\zeta=10^{-9}$.}\label{vsR1andVsPu}
%\end{minipage}\;
%\begin{minipage}[htb]{0.24\linewidth}
%\centering\includegraphics[width=\linewidth]{figsF5/conf_vsPu_vsR1_detect.eps}
%\caption{The probability of successful detection or decoding vs. $R_1$ and $P_u$, using $\theta_u=-40$ dB, $\theta_b=-60$ dB and $\zeta=10^{-9}$.}\label{vsPuVsR1}
%\end{minipage}\;
%\begin{minipage}[htb]{0.24\linewidth}
%\centering\includegraphics[width=\linewidth]{figsF5/conf_vsZeta.eps}
%\caption{The probability of successful detection or decoding vs. $\zeta$ using $R_1=7v$, $\theta_u=\theta_b=-40$ dB.}\label{vsZeta}
%\end{minipage}
%\vspace{-0.3cm}
%\end{figure*}

Fig. \ref{vsR1} plots the decoding and detection probabilities vs. $R_1$. As anticipated, the probability of decoding at the UE and decoding $1^{st}$ at the BS is unaffected by $R_1$, while the probability of detecting $1^{st}$ or $2^{nd}$ and decoding $2^{nd}$ all decrease with $R_1$. When intercell interference is taken into account, we observe that at smaller $R_1$ detecting $1^{st}$ is superior to detecting $2^{nd}$, while at larger $R_1$ detecting $2^{nd}$ is better as removing intracell interference compensates for the lower signal strength. Since decoding $1^{st}$ is always superior to decoding $2^{nd}$, at larger $R_1$ this SuIC order is optimum. At smaller $R_1$, decoding $2^{nd}$ is worse than decoding $1^{st}$, but the gap is not very large. This highlights that unless there are stringent requirements on the uplink decoding probability, below a certain $R_1$ threshold, it is better to detect $1^{st}$ and then decode, while after this threshold, decoding $1^{st}$ and then detecting is superior. Without considering intercell interference, in addition to overestimation of performance, it appears that decoding $1^{st}$ is always the optimum SuIC order which is misleading. This highlights the importance of studying a large network.

Fig. \ref{vsPu} plots the decoding and detection probabilities vs. $P_u$. When intercell interference is considered, the probability of decoding $1^{st}$ increases with $P_u$ as the communication signal strength grows but eventually saturates as the impact of interefering UEs also grows. As detecting $2^{nd}$ depends on decoding $1^{st}$, a similar trend is seen and at high $P_u$ the high interference from UEs causes it to decrease slowly. The probability of detecting $1^{st}$ decreases with $P_u$ as the interference from UEs and intracell interference from the decoding signal grows. Decoding $2^{nd}$ at the BS initially grows with $P_u$ due to the signal component growing; however, the impact of increasing interference combined with the need to successfully detect $1^{st}$ results in the probability decreasing with $P_u$ at higher $P_u$. We again observe a cross-over value of $P_u$ before which detecting $1^{st}$ is superior and after which detecting $2^{nd}$ is superior. When intercell interference is ignored, some of the trends are similar: the probability of decoding $1^{st}$ grows with $P_u$ as the signal strength grows. Detecting $2^{nd}$ grows as it depends on decoding $1^{st}$. The probability of detecting $1^{st}$ decreases with $P_u$ due to the growing intracell interference from the decoding signal. Decoding $2^{nd}$ has a tradeoff because the signal strength is growing but it also depends on detecting $1^{st}$ which is decreasing. Similar to Fig. \ref{vsR1}, not considering intercell interference eludes that decoding $1^{st}$ is always optimum, which is misleading for a real network.

In Fig. \ref{vsPuVsR1}, we observe that the threshold distance $R_1$ after which detecting $2^{nd}$ is superior to detecting $1^{st}$ decreases with $P_u$. Similarly, the threshold $P_u$ after which detecting $2^{nd}$ is superior decreases with $R_1$. These results shed light on carefully selecting the SuIC order in our setup that results in overall superior performance for both detection and decoding.% at the BS. 

Fig. \ref{vsZeta} plots the decoding and detection probabilities vs. $\zeta$. As anticipated, they decrease with $\zeta$. This happens at lower values for detection due to the higher vulnerability attributed to the double path loss in the signal component. Since decoding $2^{nd}$ only happens after successful detection $1^{st}$, this too is more vulnerable. Decoding $1^{st}$ at the BS is significantly more resilient. Decoding at the UE is even more resilient because the BS's transmit power is higher than the UE's.

%\newpage

\section{Conclusion}
To efficiently reuse the spectrum, we study a large FD cellular network with ISAC. The BS's downlink communication signal is also used to probe the target. Monostatic detection is considered; thus at the BS, two signals are received: the communication-mode uplink signal to be decoded and the radar-mode signal to be detected. After SIC, SuIC is a natural approach to decode and detect at the BS; however, SuIC requires ordering of signals based on channel strength which is not available as the target is unknown. The radar-mode signal suffers a double path-loss which eludes that the radar-mode signal is weaker, however, the uplink communication signal is sent with $P_u$ which is generally lower than $P_b$. Further, intercell interference plays a large role on both SINRs reducing the disparity between them. The optimum order for SuIC is thus not clear. We investigate both ordering schemes analytically. Our results highlight the significance of careful SuIC ordering. We find that for targets within a threshold distance from the BS, detecting $1^{st}$ is superior and decoding $2^{nd}$ results in only a small uplink performance degradation. Beyond this distance, decoding $1^{st}$ and detecting $2^{nd}$ are both superior. Further, the threshold distance decreases with $P_u$. We show that the optimum SuIC order also changes beyond a threshold $P_u$. We find that when intercell interference is ignored, decoding $1^{st}$ and detecting $2^{nd}$ appears to always be optimum, which is misleading. We highlight the vulnerability of each mode for both SuIC orders with increasing RSI.

%%\section*{Acknowledgment}
%%
%%
%%The authors would like to thank...



%\bibliographystyle{IEEEtran}
%%%\begin{thebibliography}{1}
%%%
%%%\bibitem{IEEEhowto:kopka}
%%%H.~Kopka and P.~W. Daly, \emph{A Guide to \LaTeX}, 3rd~ed.\hskip 1em plus
%%%  0.5em minus 0.4em\relax Harlow, England: Addison-Wesley, 1999.
%%%
%%%\end{thebibliography}

%%\bibliographystyle{unsrt}

\bibliographystyle{IEEEtran}
\bibliography{References}

\end{document}